\documentclass[12pt]{article}
\usepackage{amssymb}
\usepackage{amsmath}
\def\dd{{\rm d}}\def\ii{{\rm i}}
\def\vd{\vec{\dd}}

\def\beq{\begin{equation}}\def\eeq{\end{equation}}
\def\bea{\begin{eqnarray}}\def\eea{\end{eqnarray}}

\begin{document}

\title{Novel definition of Grassmann numbers and spinor fields}

\author{Roman Sverdlov
\\Physics Department, University of Michigan,
\\450 Church Street, Ann Arbor, MI 48109-1040, USA}
\date{August 14, 2008}
\maketitle

\begin{abstract}

\noindent 

The goal of this paper is to define fermionic field in terms of non-orthonormal vierbeins, where fluctuations away from orthonormality are viewed as fermionic field. Furthermore, Grassmann numbers are defined in a way that makes literal sense. 

\end{abstract}

\noindent{\bf 1. Introduction}

The goal of this paper is to come up with more intuitive definition of fermions. By intuitive I mean that adresses two of the following issues:

1)How to define Grassmann numbers and their integrals in such a way that they make literal sense.

2)How to define spinors in such a way that I won't have to appeal to notions such as "spin up" and "spin down" which seem to single out $z$ axis as "better" than the other two axes. 

In the first section of this paper we will define Grassmann numbers in a way that they are defined as individual elements of the set outside the integration. I have also made sure that the integration makes literal sense for arbitrary functions that don't have to be expressed in algebraic way, and it happens to coincide with a desired results for the commutting numbers. The key to doint that is to make sure that the space of Grassmann numbers is equipped both with the commutting dot product and anticommuting wedge product, so that, for example, $\int d \theta_1 d \theta_2 \theta_1 \theta_2$ becomes $\int (\vec{d} \theta_1 \wedge d \theta_2) \cdot (\theta_1 \wedge \theta_2)$

In the next section, I move on to part 2. The latter was already adressed in Ref \cite{paper2} where I have assumed a toy model that there are no Grassmann numbers. In that paper, I have gotten rid of unwanted fermionic degrees of freedom by trading off fermionic degrees of freedom with vierbein ones, while appealing to the Lorentz symmetry that mixes the two. However, that argument no longer works in case of the situation where the fermionic fields are Grassmannian since vierbeins are not. Therefore, this paper takes a different approach. Instead of identifying fermionic field with vierbines, it identifies it with FLUCTUATION of vierbines away from their orthonormal state. That is, vierbines are replaced by vectors that are no longer assumed to be orthonormal. This means they have $16=6+6+4$ degrees of freedom: 

(i)$6$ degrees of freedom associated with their "projection" into a space of orthonormal vectors

(ii) 6 degrees of freedom associated with fluctuation away from orthogonality of their norm 1 components 

(iii) 4 degrees of freedom associated with norm of each. 

In these paper they are re-interpretted as follows: (i) remain to be vierbeins, (ii) are now interpretted as 6 out of 8 needed fermionic degrees of freedom. The remaining 2 fermionic degrees of freedom are either added by hand or else are borrowed from (iii). In the latter case, the remaining two degrees of freedom of (iii) can be viewed as the two degrees of freedom of complex charge scalar field in which case they can be interpretted as superpartners of my fermion, or else they can be constructed as anticommutting and be used as Fadeev Popov ghosts for some gauge interaction.

\bigskip
\noindent{\bf 2. Literal Interpretation of Grassmann Numbers}

\noindent My goal is to view Grassman numbers as elements of vector space, S,
equipped both with commutting dot product ( $\cdot$ ) , anticommuting wedge product ($\wedge$), and measure $\xi$. I would like my integration to be well defined for any function $\vec{F} \colon S \rightarrow S \oplus (S \wedge S) \oplus (S \wedge S \wedge S) \oplus \ldots$ where $S \wedge S$ consists of elements of the form $a \wedge b$ where $a \in S$ and $b \in S$, $S \wedge S \wedge S$ consists of elements of the form $a \wedge b \wedge c$ where $a$ , $b$, and $c$ are elements of $S$, etc. We would like our integral to be of the form
\beq
\int (\vec{d}_{\xi} x_1 \wedge \vec{d}_{\xi} x_2 \ldots \wedge \vec{d}_\xi x_n) \cdot \vec{F}(x_1, \ldots, x_n)\;, \label{integral}
\eeq
where $\vec{d}_{\xi} x_k = \xi(x_k) \hat{x}_k\, \dd x_k$ with $\xi(x_k)$ being a measure, whose values can be both positive and negative, $\hat{x}_k$ being unit vector in the $x_k$ direction; and $\vec{x}_k = x_k \hat{x}_k$.

Thus, our definition of integral is intended to work for all functions $\vec{F}$ , not neceserely linear ones. Furthermore, the definition of integral is independent of our ability to express $\vec{F}$ in algebraic form. This allows us to view Grassmann integration is literal. 

Of course, in order to above integration to be considered Grassmann, a certain conditions need to be met: If we let $\vec{\dd}_{\xi} x = \xi(x)\, \hat{x}\, \dd x$ and $\vec{x}
= x\, \hat{x}$, where $\hat{x}$ is a unit vector in the $x$ direction, then

\bea
& &\int \vd_{\xi}x \cdot \vec{x} = \int (\vd_{\xi} x \wedge 
\vd_{\xi} y) \cdot (\vec{x} \wedge \vec{y}) =1
\\
& &\int \vd_{\xi}x = \int \vd_{\xi}x \wedge 1
= \int (\vd_{\xi}x \wedge \vd_{\xi} y) \cdot \vec{x} = 0
\\ 
& &\int (\vd_{\xi} x\, \wedge \vd_{\xi} y) \cdot \vec{f} (x,y) = \int 
\vd_{\xi} x \cdot \bigg(\int \vd_{\xi} y \cdot \vec{f}(x,y) \bigg)\,.
\eea
The first two of the above equations are what we expect of Grassmann 
variables. The last equation doesn't make sense in terms of standard 
Grassmann theory, since in order to say that $\int \dd\theta_1\, \dd\theta_2\,
\theta_1\, \theta_2 = \int \dd\theta_1\, (\int \dd\theta_2\, \theta_1\,
\theta_2)$ we need to define $\int \dd\theta_2\, \theta_1\, \theta_2$,
which we can't do since its value would be Grassman number whose definition is unavailable in standard theory. But this would be one of the aspects that I intend to 
change: since I would like Grassmann numbers, on their own, to make 
literal sense, I would also like integrals such as above to make 
literal sense as well.

The way we would approach it is to pretend that we have a definition 
of dot and wedge products, which we don't. Thus, we would evaluate 
above integrals in terms of un-computed wedge and dot products. Since 
we know what we expect these integrals to be, this would tell us what 
we expect wedge and dot products to be, as well. 

From the requirement that
\beq
0 = \int \vec{\dd}_{\xi}\,x = \int \dd x\, \xi(x)\, \hat{x}
= \hat{x} \int \xi(x)\, \dd x\;,
\eeq 
we see that 
\beq
\int \xi(x)\, \dd x = 0\;;
\eeq
in other words, unlike what we are used to, the measure has both positive 
and negative values. 

From the requirement that
\beq
1 = \int \vd_{\xi}x \cdot \vec{x} = \int (\dd x\, \xi(x) \hat{x}) \cdot 
(\hat{x} x) = \hat{x} \cdot \hat{x} \int x\, \xi(x)\, \dd x\;,
\eeq
we obtain a condition which can be satified by setting
\beq
\hat{x} \cdot \hat{x} = 1\;,\qquad
\int x\, \xi(x)\, \dd x = 1\;.
\eeq
Now let us move to the multiple-integral example:
\bea
& &1 = \int \vd_{\xi} x \cdot \bigg(\int \vd_{\xi} y \cdot
(\vec{x} \wedge \vec{y} )\bigg)
= \int \bigg[\dd x\, \xi(x)\, \hat{x} \cdot \bigg(\int \dd y\, \xi(y)\,
\hat{y} \cdot (xy\, \hat{x} \wedge \hat{y})\bigg)\bigg] \nonumber\\
& &\kern9pt = \hat{x} \cdot(\hat{y} \cdot (\hat{x} \wedge \hat{y}))
\bigg(\int x\, \xi(x)\, \dd x \bigg) \bigg(\int y\, \xi(y)\, \dd y\bigg).
\eea
Since we have already established that
\beq
\int x\, \xi(x)\, \dd x = \int y\, \xi(y)\, \dd y = 1\;,
\eeq
the above calculation tells us that
\beq
\hat{x} \cdot (\hat{y} \cdot (\hat{x} \wedge \hat{y})) = 1\;,
\eeq
which can be accomplished by setting
\beq
\hat{y} \cdot (\hat{x} \wedge \hat{y}) = \hat{x}\;.
\eeq
By a similar argument we can show that
\beq
(\hat{y} \wedge \hat{z}) \cdot (\hat{x} \wedge \hat{y}
\wedge \hat{z}) = \hat{x}
\eeq
and
\beq
\hat{z} \cdot (\hat{x} \wedge \hat{y} \wedge \hat{z})
= \hat{y} \wedge \hat{z}\;.
\eeq
However, this relationship makes it a little more tricky to define the
dot product consistently, due to the anticommutativity of $\wedge$:
\beq
\hat{y} \cdot (\hat{y} \wedge \hat{x})
= -\hat{y} \cdot (\hat{x} \wedge \hat{y})=-\hat{x}\;.
\eeq
The way I will handle it is by associating unit vectors with elements 
of totally ordered set, thus making a default decision between 
$\hat{x} \wedge \hat{y}$ versus $\hat{y} \wedge \hat{x}$. I will then 
use the power of $-1$ to extend my definition of wedge product to the 
reverse orders. More precisely, I will associate vectors with 
functions on totally ordered set $S= \lbrace s_1, s_2, \ldots, s_n 
\rbrace$ For simplicity, I will define ordering in such a way that 
$s_i < s_j$ if and only if $ib \rbrace$.
The dot product, on the other hand, will be defined in the following way:

\noindent{\em Definition:\/} Let $p_1$ and $p_2$ be two polynomials
over $S$. Then $p_1 \cdot p_2$ is another polynomial over $S$ such
that for every $T \subset S$,
\beq
(p_1 \cdot p_2)(T) = \sum_{(U \setminus V) \cup (V \setminus U) = T} 
p_1(U)\, p_2(V)\;.
\eeq
Finally, in order to have definition of the derivative, we need 
definition of ratio. I will make analogy with the set of integers 
where ratio is not everywhere defined and claim that the same is okay 
here. Thus, I will make the following definition:

\noindent{\em Definition:\/} Let $\vec{a}$ and $\vec{b}$ be two Grassmann polynomials. If there exists a Grassmann polynomial $\vec{c}$ such that
$\vec{a} \wedge \vec{c} = \vec{b}$ then we say that $\vec{c}= 
\vec{b}/\vec{a}$. If such $\vec{c}$ doesn't exist, then 
$\vec{b}/\vec{a}$ is not well defined.

The important thing is that the fraction was defined in terms of the wedge 
product, as opposed to the dot product, and also that the wedge product was 
ordered in the way it was. This would allow us to define derivatives 
in the way we expect them to be.

\bigskip
\noindent{\bf 3 Spinor field as part of geometry}
 $$
\noindent 
$$

As explaned in  Ref \cite{paper 2} In the toy model where fermionic fields are complex valued as opposed to Grassmanian, we notice an interesting feature: spinor has $4$ complex degrees of freedom, which means $8$ real degrees of freedom. At the same time, the number of degrees of freedom associated with choice of vierbeins is $6$. Thus, the total number of degrees of freedom is $8-6=2$ . This means that we can trade spinor degrees of freedom with vierbein ones by always selecting frame in which spinor takes a form 

\bea u=\left( \begin{array}{ccc}
\chi_p \\
0 \\
\chi_a \\
0 
\end{array} \right) \eea

where $\chi_p$ and $\chi_a$ correspond to particle and antiparticle amplitudes, and are both real. Thus, we can describe spinor field completely in terms of scalar fields $\chi_1$ and $\chi_2$ and four orthonormal vector fields that are interpretted as veirbines and determine local frame.

The obvious obstacle to the above is the fact that spinor fields are grassmanian while vierbeins are real. Of course, the fact that we have interpretted Grassmann variables in terms of real numbers somewhat alliviates the situation, but not completely: we have to define $\xi (\psi_i)$ as well as $\hat{\psi_i}$ for all values of $i$. At the same time, neither of these are functions of vierbeins since the latter are not viewed as Grassmann. Hence, these don't respect the rotational symmetry I depend upon in my argument. Thus, if we insist on geometry, we would have to introduce TWO separate frames. One frame would give us vierbeins that are no longer viewed as part of the definition of spinor field, while the other frame will determine the spinor field -- namely, the latter would be given by "rotating" $(\chi_p , 0, \chi_a , 0)$ from one of these two frames to the other. 

While the above can be done, this ruins the beauty that comes out from counting degrees of freedom. The way to restore that beauty is to make sure that our second frame can be obtained from the first frame, hence the only TRUE degrees of freedom are the ones corresponding to the latter. This can be done by the following trick: we notice that since vierbeins are part of the field, we would like to view them as fields. This means that they are not neceserely orthonormal. Instead of restricting them to being orthonormal, we will let them be whatever they happen to be, so we will replace $e_0^{\mu}$, $e_1^{\mu}$, $e_2^{\mu}$ and $e_3^{\mu}$ by  $A^{\mu}$, $B^{\mu}$, $C^{\mu}$ and $D^{\mu}$ respectively. Then we will use Gramm Schmidt process to enforce orthonormality. This will give us the two frames that we are looking for: one is the original non-orthonormal frame, and the other is the one obtained from original one by Gramm Schmidt process: 

\bea  e_0^{\mu} (A) = \frac{A^{\mu}}{\sqrt{A^{\nu}A_{\nu}}} \eea

\bea e_1^{\mu} (A, B) = \frac{B^{\mu} - e_0^{\nu}B_{\nu} e_0^{\mu}}{\sqrt{(B^{\rho} - e_0^{\alpha}B_{\alpha} e_0^{\rho})(B_{\rho} - e_0^{\beta}B_{\beta} e_{0 \rho})}} \eea

\bea e_2^{\mu} (A, B, C) = \frac{C^{\mu} - e_0^{\alpha}C_{\alpha} e_0^{\mu}-e_1^{\beta}C_{\beta} e_1^{\mu}}{\sqrt{(C^{\rho} - e_0^{\gamma}C_{\gamma} e_0^{\rho}- e_1^{\delta}C_{\delta} e_1^{\delta})(C_{\rho} - e_0^{\delta}C_{\delta} e_{0 \rho}- e_1^{\epsilon}C_{\epsilon} e_{1 \rho})}} \eea

\bea e_3^{\mu} (A, B, C, D) = \frac{D^{\mu} - e_0^{\alpha}D_{\alpha} e_0^{\mu}-e_1^{\beta}D_{\beta} e_1^{\mu}-e_2^{\gamma}D_{\gamma} e_2^{\mu}}{\sqrt{(D^{\rho} - e_0^{\delta}D_{\delta} e_0^{\rho}- e_1^{\epsilon}D_{\epsilon} e_1^{\rho}-e_2^{\phi}D_{\phi} e_2^{\rho})(D^{\rho} - e_0^{\chi}D_{\chi} e_0^{\rho}- e_1^{\eta}D_{\eta} e_1^{\rho}-e_2^{\xi}D_{\xi} e_2^{\rho})}} \eea

This will give us the definition of spinor field: if we let 

\bea f_0^{\mu}=\frac{A^{\mu}}{\sqrt{A^{\nu}A_{\nu}}} , \; f_1^{\mu}=\frac{B^{\mu}}{\sqrt{B^{\nu}B_{\nu}}} ,  \; f_2^{\mu}=\frac{C^{\mu}}{\sqrt{C^{\nu}C_{\nu}}} ,  \;f_3^{\mu}=\frac{D^{\mu}}{\sqrt{D^{\nu}D_{\nu}}}\eea
   
   then we can define our spinor to be

\bea
& & \psi_i(\chi_p , \chi_a , \frac{A}{\vert A \vert}, \frac{B}{\vert B \vert}, \frac{C}{\vert C \vert}, \frac{D}{\vert D \vert}) \\
& & =\big(\exp\big\{ -{\textstyle\frac{\ii}{4}} (\ln (e^{-1} (A,B,C,D) f (A,B,C,D)))_{\mu \nu} \sigma^{\mu\nu} \big\}\, \big)_{ij} \big(\chi_p \delta^j_1 + \chi_a \delta^j_3 \big) \;. 
\nonumber
\eea

After having done that, we will take advantage of the fact that we have no information about $\xi$ function other than the two integrals, which gives us freedom to define it to be derivative of delta function, 

\bea \xi (x) = \frac{d \delta (x)}{dx} \eea

where our delta function does have finite width, albeit very small, which might be expressed by replacing delta function with $\sqrt{\frac{a}{\pi}} e^{-ax^2}$ which gives us  

\bea \xi (x) = \frac{a^{1.5}}{a^{0.5}} e^{-ax^2} \eea

for some very large $a$. It is easy to see that this satisfies both of the desired properties for $\xi$ and also it would assure us that $A$, $B$, $C$ and $D$ are approximately orthonormal, even though not exactly. This would save us from worrying about some of the global issues in spinor transformations, such as the fact that Lorentz group has two connected components rather than one. 

Now it is time to move to integration. Since we intend to view $A$, $B$, $C$, $D$ as physical fields, I would like to integrate over them. This means that I have to replace the measure $\xi$ on the $\psi$ space with a measure $\lambda$ on $\chi_p \chi_a ABCD$ space. This can be done as follows:

\bea & & \lambda (\chi_p , \chi_a , A, B, C, D) =\xi \Big[\psi \Big(\chi_p, \chi_a, \frac{A}{\vert A \vert}, \frac{B}{\vert B \vert}, \frac{C}{\vert C \vert}, \frac{D}{\vert D \vert}\Big)\Big] \times \\ \nonumber
& &  \times lim_{\epsilon \rightarrow 0} \epsilon \mu^{-1} \Big\{ \chi_p ' , \chi_a ', A' , B' , C' , D' \Big\vert \big\vert \psi\Big(\chi_p ', \chi_a ', \frac{A'}{\vert A' \vert}, \frac{B'}{\vert B' \vert},\frac{C'}{\vert C' \vert},\frac{D'}{\vert D' \vert} \Big) \nonumber \\
& & - \psi \Big(\chi_p , \chi_a , \frac{A}{\vert A \vert},\frac{B}{\vert B \vert}, \frac{C}{\vert C \vert} , \frac{D}{\vert D \vert} \Big) \big\vert < \epsilon  \wedge \forall a \Big[\Big(e_a^{\mu} \Big( \frac{A'}{\vert A' \vert},\frac{B'}{\vert B' \vert}, \frac{C'}{\vert C' \vert}, \frac{D'}{\vert D' \vert}\Big) - e_a^{\mu} \Big( \frac{A}{\vert A \vert},\frac{B}{\vert B \vert},\frac{C}{\vert C \vert},\frac{D}{\vert D \vert}\Big) \Big)  \nonumber \\
& & \Big(e_{a\mu} \Big(\frac{A'}{\vert A' \vert}, \frac{B'}{\vert B' \vert},\frac{C'}{\vert C' \vert}, \frac{D'}{\vert D' \vert}\Big)   - e_{a\mu} \Big(\frac{A'}{\vert A' \vert}, \frac{B'}{\vert B' \vert},\frac{C'}{\vert C' \vert}, \frac{D'}{\vert D' \vert} \Big)\Big) \Big] < \epsilon^2 \Big\}  \eea 

where $\mu$ is a usual measure on Eucledian space $\mathbb{R}^{18}$ . Strictly speaking, due to the fact that spacetime is not compact, that measure is not well defined. But this can eaasilly be dealt with if we impose restrictions

\bea A^{\mu}A_{\mu} + B^{\mu}B_{\mu} + C^{\mu}C_{\mu} + D^{\mu}D_{\mu}  \leq r^2 \eea
\bea \vert \chi_p \vert \leq r \wedge \vert \chi_a \vert \leq r \eea

for some large $r$ .  Just to remind the reader, since Grassmann numbers are defined in terms of real numbers, in above expression $\chi_p$ and $\chi_a$ are real, since we haven't multiplied them by anticommuting unit vectors yet, hence their squares are non-zero, which means that absolute value is well defined.

 Even though us having plus signs instead of minus signs in above equation might appear to violates relativity, we don't have to worry about that because $A$, $B$, $C$, and $D$ are distinct from vierbeins which means they are interpretted as fields as opposed to reference frame. 

Now that we have gotten rid of $\xi (\psi_a)$ we also have to get rid of $\hat{\psi}_a$ . That we do by simply replacing real and imaginary parts of $\hat{\psi_1}$, $\hat{\psi_2}$, $\hat{\psi_3}$,  and $\hat{\psi_4}$ with $\hat{r_1}$ through $\hat{r_8}$. The latter are viewed as constant unit vectors, and are no longer interpretted as part of any field. We then define $\vec{\phi}$ as follows:

\bea & & \vec{\psi} (\chi_p, \chi_a, A,B,C,D) = \hat{r_1} Re (\psi_1 (\chi_p, \chi_a, A, B, C,D)) + i \hat{r_2} Im (\psi_1 (\chi_p, \chi_a, A, B, C,D)) \nonumber \\
& & + \hat{r_3} Re(\psi_2 (\chi_p, \chi_a, A,B,C,D))+  i \hat{r_4} Im(\psi_2 (\chi_p, \chi_a, A,B,C,D)) \nonumber \\
& & + \hat{r_5} Re(\psi_3 (\chi_p, \chi_a, A,B,C,D)) + i \hat{r_6} Im (\psi_3 (\chi_p, \chi_a, A,B,C,D)) \nonumber \\
& & + \hat{r_7} Re (\psi_4 (\chi_p, \chi_a, A,B,C,D)) + i \hat{r_8} Im (\psi_4 (\chi_p, \chi_a, A,B,C,D)) \eea

Then our Grassmannian integral becomes

\bea & & Z = \int d^d A d^d B d^d C d^d D d \chi_p d \chi_a \; \lambda (\psi_1 (\chi_p, \chi_a, A,B,C,D))  \lambda (\psi_2 (\chi_p, \chi_a, A,B,C,D)) \nonumber \\
& & \times \lambda (\psi_3 (\chi_p, \chi_a, A,B,C,D)) \lambda (\psi_4 (\chi_p, \chi_a, A,B,C,D))  (\hat{r_1} \wedge . . . \wedge \hat{r_8}) \cdot e^{iS (\vec{\psi} (\chi_p, \chi_a, A,B,C,D))}  \eea

\bigskip
\noindent{\bf 4. Taking advantage of norm degrees of freedom}
 $$
\noindent 
$$

\bigskip
\noindent{\bf 4.1\;    First pair of degrees of freedom}
 $$
\noindent 
$$

In the previous section, we have illustrated a way of defining fermions by using scalar fields $\chi_p$ and $\chi_a$ together with the degrees of freedom associated with fluctuations of vierbein-like vector fields away from their orthogonal position. However, we have gotten rid of the four degrees of freedom associated with their fluctuations away from norm $1$. This suggests that we can get rid of $\chi_p$ and $\chi_a$ in favor of two of these four degrees of freedom. This means we have to introduce functions $\chi_p (A, B, C, D)$ and $\chi_a (A, B, C, D)$. Since all the earlier results were independent of magnitude, regardless what our two functions are, they would effectively amount to introducing two out of four magnitude degrees of freedom.  We then rewrite $\psi$ as only a function of $A$, $B$, $C$ and $D$:

\bea \psi_a (A, B, C, D ) = \psi_a \Big(\chi_p (A,B,C,D) , \chi_a (A, B, C, D) , \frac{A}{\vert A \vert}, \frac{B}{\vert B \vert}, \frac{C}{\vert C \vert}, \frac{D}{\vert D \vert} \Big) \eea

Likewise, 

\bea & & \vec{\psi} (A,B,C,D) = \hat{r_1} Re (\psi_1 (\chi_p (A,B,C,D) , \chi_a (A,B,C,D) , A, B, C,D)) \nonumber \\ 
& & + i \hat{r_2} Im (\psi_1 (\chi_p (A, B, C, D), \chi_a (A, B, C, D), A, B, C,D)) \nonumber \\
& & + \hat{r_3} Re(\psi_2 (\chi_p (A, B, C, D), \chi_a (A, B, C, D), A,B,C,D)) \nonumber \\ 
& & +  i \hat{r_4} Im(\psi_2 (\chi_p (A, B, C, D) , \chi_a (A, B, C, D), A,B,C,D)) \nonumber \\
& & + \hat{r_5} Re(\psi_3 (\chi_p(A, B, C, D), \chi_a (A, B, C, D), A,B,C,D)) \nonumber \\
& & +i  \hat{r_6} Im (\psi_3 (\chi_p (A, B, C, D), \chi_a (A, B, C, D), A,B,C,D)) \nonumber \\
& & + \hat{r_7} Re (\psi_4 (\chi_p (A, B, C, D), \chi_a (A, B, C, D), A,B,C,D)) \nonumber \\
& & + i \hat{r_8} Im (\psi_4 (\chi_p (A, B, C, D), \chi_a (A, B, C, D), A,B,C,D)) \eea

Our measure is now on the $16$ dimensional $ABCD$ space as opposed to $18$ dimensional $\chi_p \chi_a ABCD$ space, and is defined as follows: 

\bea & & \lambda (A, B, C, D) =\xi \Big[\psi \Big(\chi_p (A, B, C, D), \chi_a (A, B, C, D), \frac{A}{\vert A \vert}, \frac{B}{\vert B \vert}, \frac{C}{\vert C \vert}, \frac{D}{\vert D \vert}\Big)\Big] \times \\ \nonumber
& &  \times lim_{\epsilon \rightarrow 0} \epsilon \mu^{-1} \big( \{ A' , B' , C' , D' \big\vert \vert \psi(\chi_p (A', B', C', D'), \chi_a (A', B', C', D'), A',B',C',D') \\ \nonumber
& & - \psi (\chi_p (A,B,C,D) , \chi_a (A,B,C,D), A,B,C,D) \vert < \epsilon \wedge \\ \nonumber
& & \wedge \forall a (e_a^{\mu} ( A',B',C',D') - e_a^{\mu} ( A,B,C,D))  
(e_{a\mu} (A',B',C',D')   - e_{a\mu} (A,B,C,D)) < \epsilon^2 \} ) \eea 

Our Grassmannian integral becomes

\bea & & Z = \int d^d A d^d B d^d C d^d D \; \lambda (\psi_1 (\chi_p (A,B,C,D), \chi_a(A,B,C,D), A,B,C,D)) \\
& & \lambda (\psi_2 (\chi_p(A,B,C,D), \chi_a(A,B,C,D), A,B,C,D))  \lambda (\psi_3 (\chi_p(A,B,C,D), \chi_a(A,B,C,D), A,B,C,D)) \nonumber \\
& &  \lambda (\psi_4 (\chi_p(A,B,C,D), \chi_a (A,B,C,D), A,B,C,D))  (\hat{r_1} \wedge . . . \wedge \hat{r_8}) \cdot e^{iS (\vec{\psi} (\chi_p(A,B,C,D), \chi_a (A,B,C,D), A,B,C,D))} \nonumber \eea

\bigskip
\noindent{\bf 4.2\;  Second pair of degrees of freedom}
 $$
\noindent 
$$

Finally, we can take advantage of the two remaining degrees of freedom, $g_1 (A,B,C,D)$ and $g_2 (A, B, C, D)$ to construct scalar fields. There is no conclusive criteria of what these fields should be. So we can simply use it for our own convenience and stick them at some other, seemingly unrelated, issue where we wish there were fields but there aren't. Since in modern physics there are a lot of unresolved issues, the reader is invited to use these two remaining degrees of freedom for their own issues of interest. In this section I will present just two possibilities that are my personal favorites. They are fadeev popov ghosts and superpartners. 

It is important to stress to the reader that these two possibilities are NOT related to each other, and in fact they are probably incompatible, since fadeev popov ghosts, as they are, are used in non-supersymmetric theories. So, these two possibilities are presented as only possibilities, and should not be taken too seriously. 

\noindent{\bf  Possibile definition of fadeev popov ghosts}

One possible thing to do is to interprit these scalar fields as Fadeev Popov ghosts that are to be used in the gauge field that interacts with a fermion of our interest, thus providing an interpretation of Fadeev Popov ghosts as well. For example, Fadeev popov ghosts of weak interaction can go from the "extra components" of electron, left handed and right handed neutrino, which gives us $6$ real degrees of freedom, which matches three complex degrees of freedom of $c_a$ and $\overline{c}_a$ . In order to account for ghosts, our new measure will be 

\bea & & \lambda (A, B, C, D) = \nonumber \\
& & \xi \Big[\psi \Big(\chi_p (A, B, C, D), \chi_a (A, B, C, D), \frac{A}{\vert A \vert}, \frac{B}{\vert B \vert}, \frac{C}{\vert C \vert}, \frac{D}{\vert D \vert}\Big)\Big] \xi (g_1 (A,B,C,D)) \xi (g_2 (A,B,C,D)) \times \nonumber \\
& &  \times lim_{\epsilon \rightarrow 0} \epsilon \mu^{-1} \big \{ A' , B' , C' , D' \big\vert \vert \psi(\chi_p (A', B', C', D'), \chi_a (A', B', C', D'), A',B',C',D') \nonumber \\
& & - \psi (\chi_p (A,B,C,D) , \chi_a (A,B,C,D), A,B,C,D) \vert < \epsilon \wedge \\ \nonumber
& & \wedge \forall a (e_a^{\mu} ( A',B',C',D') - e_a^{\mu} ( A,B,C,D))  
(e_{a\mu} (A',B',C',D')   - e_{a\mu} (A,B,C,D)) < \epsilon^2 \nonumber \\
& & \wedge \vert g_1 (A',B',C',D') - g_1 (A,B,C,D) \vert \leq \epsilon  \wedge \vert g_2 (A',B',C',D') - g_2 (A,B,C,D) \vert \leq \epsilon\} \eea 

The integral will be 

\bea & & Z = \int d^d A d^d B d^d C d^d D \; \lambda (\psi_1 (\chi_p (A,B,C,D), \chi_a(A,B,C,D), A,B,C,D)) \\
& & \lambda (\psi_2 (\chi_p(A,B,C,D), \chi_a(A,B,C,D), A,B,C,D))  \lambda (\psi_3 (\chi_p(A,B,C,D), \chi_a(A,B,C,D), A,B,C,D)) \nonumber \\
& &  \lambda (\psi_4 (\chi_p(A,B,C,D), \chi_a (A,B,C,D), A,B,C,D))  \times \nonumber \\
& & \times (\hat{s_1} \wedge \hat{s_2} \wedge \hat{r_1} \wedge . . . \wedge \hat{r_8}) \cdot \exp \big( iS (\vec{\psi} (\chi_p(A,B,C,D), \chi_a (A,B,C,D), A,B,C,D)) \nonumber \\
& & + iS_g (\hat{s_1} g_1(A,B,C,D) + i \hat{s_2} g_2 (A,B,C,D)) \big) \eea

where $\hat{s_1}$ and $\hat{s_2}$ are unit vectors introduced for the ghost fields. 

\noindent{\bf  Possibile model of superpartners}

We now explore the other possibility of using the two remaining degrees of freedom. Of course, in order to use them in any way other than the way we have just used them, we have to abandon our model of ghosts, in order to get these two degrees of freedom back. Thus, before proceeding, a reader should fully realize that the two models are unrelated and incompatible. Appart from the fact that the same couple of degrees of freedom can not be used twice, fadeev popov ghosts, as we know them, are part of non-supersymmetric models. With this in mind, let us proceed with an alternative model.

We can interpret them as usual commutting bosonic fields.  In this case we get rid of $s_1$, $s_2$, $\xi (g_1)$ and $\xi (g_2)$ which gives us the following: 

\bea & & \lambda (A, B, C, D) = \xi \Big[\psi \Big(\chi_p (A, B, C, D), \chi_a (A, B, C, D), \frac{A}{\vert A \vert}, \frac{B}{\vert B \vert}, \frac{C}{\vert C \vert}, \frac{D}{\vert D \vert}\Big)\Big]  \times \nonumber \\
& &  \times lim_{\epsilon \rightarrow 0} \epsilon \mu^{-1} \big \{ A' , B' , C' , D' \big\vert \vert \psi(\chi_p (A', B', C', D'), \chi_a (A', B', C', D'), A',B',C',D') \nonumber \\
& & - \psi (\chi_p (A,B,C,D) , \chi_a (A,B,C,D), A,B,C,D) \vert < \epsilon \wedge \\ \nonumber
& & \wedge \forall a (e_a^{\mu} ( A',B',C',D') - e_a^{\mu} ( A,B,C,D))  
(e_{a\mu} (A',B',C',D')   - e_{a\mu} (A,B,C,D)) < \epsilon^2 \nonumber \\
& & \wedge \vert \phi_1 (A',B',C',D') - \phi_1 (A,B,C,D) \vert \leq \epsilon  \wedge \vert \phi_2 (A',B',C',D') - \phi_2 (A,B,C,D) \vert \leq \epsilon\} \eea 

and

\bea & & Z = \int d^d A d^d B d^d C d^d D \; \lambda (\psi_1 (\chi_p (A,B,C,D), \chi_a(A,B,C,D), A,B,C,D)) \\
& & \lambda (\psi_2 (\chi_p(A,B,C,D), \chi_a(A,B,C,D), A,B,C,D))  \lambda (\psi_3 (\chi_p(A,B,C,D), \chi_a(A,B,C,D), A,B,C,D)) \nonumber \\
& &  \lambda (\psi_4 (\chi_p(A,B,C,D), \chi_a (A,B,C,D), A,B,C,D))  \times \nonumber \\
& &\times ( \hat{r_1} \wedge . . . \wedge \hat{r_8}) \cdot \exp \big( iS (\vec{\psi} (\chi_p(A,B,C,D), \chi_a (A,B,C,D), A,B,C,D)) \nonumber \\
& & + iS_g (\phi_1(A,B,C,D) + i \phi_2 (A,B,C,D)) \big) \nonumber \eea

The commutting case makes it very tempting to consider them to be superpartners of a given particle. While it is certainly an interesting possibility to explore, it is important to realize that while they are "partner" they don't have to be superpartners. There is no obvious symmetry that relates them to the fermions, although we can always engeneer that symmetry by tempering with measure. Hence, we have no reason to expect them to even have the same mass as fermions. So it is possible that, for example, a "partner" of electron is Higgs boson that has nothing to do with electron. However, it is certainly possible to adjust things to "manufacture" supersymmetry by simply constructing Lagrangian that is written in non-supersymmetric form in regular coordinates (as opposed to superspace), but simply happens to satisfy supersymmetry transformations. In light of the fact that in this paper we have already "manufactured" non-supersymmetric theory by adjusting measure, there is no reason to stop us from "manufacturing" supersymmetric theories as well if we want to. 

Ironically, while the definition of Grassmann numbers allows me to define superspace in a literal form, doing the latter would not go together with the above model of superpartners. After all, if we do define superspace in a standard way, we would need to introduce a superfield which will probably be separate scalar field satisfying some constraints. Since spinor components that define superspace will now be part of the trajectory, while the scalar superfield will be a field on the space of these trajectories, they will no longer be allowed to mix. But, due to a familiar fact that supersymmetric theories CAN be described without the appeal to the concept of superspace, this is not an obstacle to introducing supersymmetric theories.

However, the superspace model itself offers a different kind of inside: since in this paper we have identified spinor with a set of four different vectors, it is the other way of saying that spinor is identified with a local frame. Thus, superspace gains a very geometric view: it is a space where point is not simply a point, but rather point plus the frame. In a sense, this is very appealing since we can't imagine a point without imagining space into which the point was placed, and the definition of space is set of coordinates. One can also think of this kind of superspace as continuum version of spin foams, which opens door to explore new set of theories. 

Thus, both definition of superspace and definition of superpartners are interesting possibilities that are, unfortunately, not compatible. But each of them is worth further exploration.

\noindent{\bf 5. Conclusion}

From what we have seen in this paper, we have found a way to introduce fermions while avoiding two of its unpleasant features: Grassmannian nature as well as inability of us to "visualize" something that has spin 1/2 rotational property. We introduce four vector fields that relate to vierbeins but don't coincide with them, we can use extra degrees of freedom in order to define fermionic fields. Furthermore, we have seen that by treating the latter degrees of freedom as elements of space equipped both with anticommutting wedge and commutting dot product, as well as measure defined in a very specific way (in particular it has both positive and negative values) then we can obtain usual quantum field theory in terms of integration defined in a literal sense of the word. This allows us to define fermions while avoiding its two unpleasant features: anticommutativity and inability to be viewed as 

Open gaps of the theory include the fact that it is non-renormalizeable. However, this obstacle can easilly be passed by since our theory is mathematically equivalent to the renormalizeable one. 

One interesting offshot of what we have found is that we can view the four separate vectors that define spinor field as local frame. This gives us a continuum version of spin foams. The important difference with this model and spin foams, however, is the fact that each spinor field has its separate set of four vectors, hence the frame is not part of geometry but rather a part of fermionic field. One can argue that this has its own appeal in a sense that if there was one set of four vectors, they might imply a "prefered frame" while in our case since each field has its own set of vectors, it is clear that there is no prefered frame other than the one that comes with a field, and this no longer appears to violate relativity any more than, say, electric charges violate translational symmetry.

\noindent{\bf Acknowledgements}
I would like to express my gratitude to Luca Bombelli, Alexi Tkachenko, James Liu, Marc Ross and Mark Galperin for useful discussions, feedback, encouragement and support.


\end{document}